# Opportunities and Challenges of Integrating ChatGPT in Education: Sentiment Analysis and Topic Modeling

Since its recent debut, ChatGPT has become a global sensation and significantly impacted the field of education. Both educational researchers and practitioners have identified opportunities as well as risks associated with the use of this novel tool in educational settings. Despite the ongoing debate, there is still no research exploring occupational differences in the perception of ChatGPT in education. In this paper, we analyzed Twitter data using topic modeling and sentiment analysis to investigate how ChatGPT is perceived and discussed differently in different occupations. Our study found diverse topics discussed including its use in schools, impact on exams, academic integrity concerns, and response accuracy evaluations. While most tweets were positive or neutral, concerns about integrity and response accuracy were evident. Analysis revealed sentiment and topic variations among users' occupations. These findings emphasize the opportunities and challenges of integrating ChatGPT in education, necessitating continued monitoring and informed policy-making for responsible utilization.

## 1 Introduction

ChatGPT has quickly become a global sensation, reaching one million users within a week and 100 million monthly active users within two months, making it the fastest-growing consumer application (Koonchanok, Pan, & Jang, 2023; Vallance, 2022). Developed by OpenAI, it is a large language model designed to follow instructions and provide detailed responses (OpenAI, 2022).

In education, ChatGPT can tailor learning experiences, develop materials, and overcome language barriers. It helps students summarize information, explain code, and provide feedback on academic tasks. However, concerns include students using ChatGPT for assignments, making it hard to detect AI-generated content, and the risk of students not improving their skills. ChatGPT can also produce misleading information, known as "hallucinations".

Despite the ongoing debate over the utilization of ChatGPT in education, scientific research and formal analysis on this topic are hardly available. Those that exist mainly investigate the sentiment and topic discussed in the social media platforms except for Gulati et al. (2024), which investigates predictors for behavioral intention to use ChatGPT. To our knowledge, there has been no prior research dedicated to exploring field of study variation in perception on ChatGPT in education. In this paper, we aim to fill this gap by providing a comprehensive analysis based on Twitter data from the perspective of occupation differences. In particular, we ask the following research questions.

(1) What topics do people discuss about ChatGPT in education?
(2) What are the general sentiments towards ChatGPT in education? What are the sentiments towards each topic?
(3) What are the occupational differences in sentiments and topics discussed in ChatGPT in education?

## 2   Related Work

Haque et al. (2022) used topic modeling and sentiment analysis on tweets, finding 52% positive, 32% negative, and 16% neutral sentiments about ChatGPT in education. Taecharungroj (2023) utilized the Latent Dirichlet Allocation (LDA) topic modeling and identified several domains where ChatGPT could effectively operate, such as creative writing, essay writing, prompt writing, code writing, and question answering. Leiter et al. (2024) found positive sentiments in scientific fields but concerns in education through emotion analysis.

Adeshola and Adepoju (2023) also used topic modeling and sentiment analysis, revealing mostly positive tweets, and further identifying six main topics. Koonchanok, Pan and Jang (2023) explored topic- and occupation-specific sentiments. Li et al. (2024) identified key concerns about ChatGPT in education, including academic integrity and skill development. Fütterer et al. (2023) found mixed sentiments on topics like cheating and opportunities from ChatGPT. Lampropoulos, Ferdig and Kaplan-Rakowski (2023) analyzed general and educational uses of ChatGPT. So et al. (2024) found neutral sentiments and eleven topics related to Generative AI in education on YouTube.

Few studies focus on public opinions about ChatGPT in education, often missing demographic perspectives. This paper addresses these gaps by using topic modeling and sentiment analysis on tweets to explore public concerns and occupational differences in sentiments and topics discussed about ChatGPT in education.

## 3   Material and methods

Figure 1. Overview of data cleaning and preprocessing.

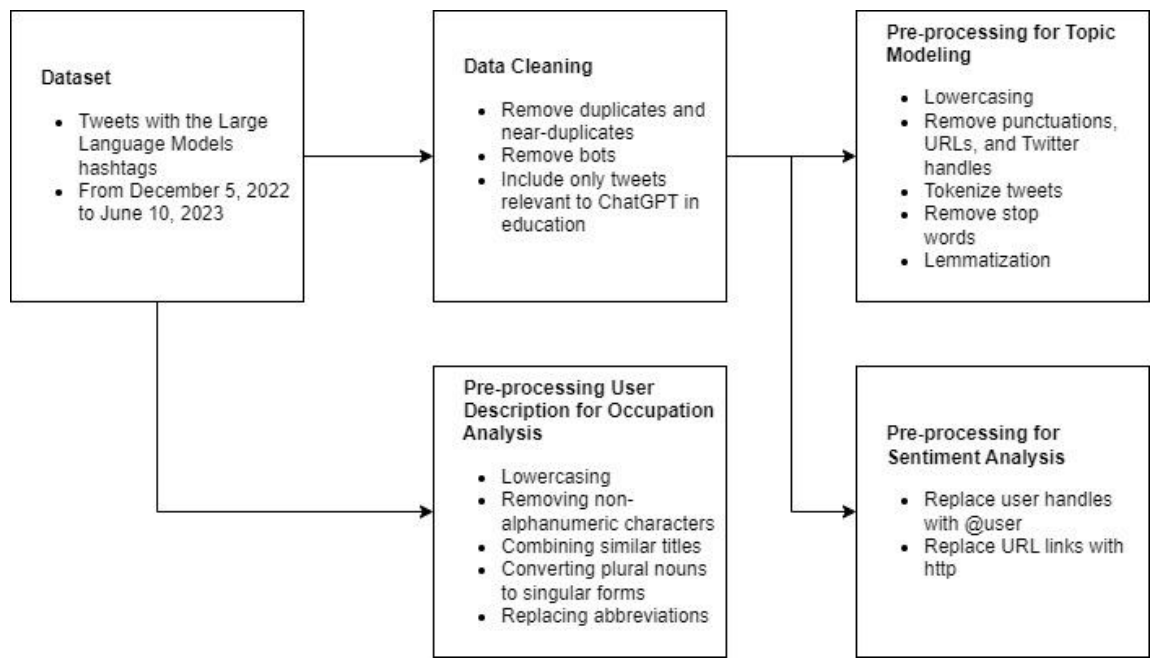

*3.1  Dataset*

Figure 1 displays the visual representation of our research framework, from acquiring the dataset to analysis methods. We utilized a dataset sourced from Kaggle[1]. This dataset contains a collection of tweets with the Large Language Models hashtags. The dataset has been continuously updated on a daily basis since December 5, 2022. We used version 172, which was last updated as of June 10, 2023.

*3.2  Data Cleaning*

First, we removed duplicates, near-duplicates, and tweets originating from automated accounts. As bots can be characterized by frequent posting and a propensity to post repeated content (Koonchanok, Pan, & Jang, 2023), we identified bots as the top 1% most active users. Detection of near-duplicates was accomplished through the implementation of MinHash, a methodology for estimating document similarity proposed by Broder (1997). To include only tweets relevant to ChatGPT in education, the tweets must include ChatGPT as a keyword, as well as keywords including words related to the educational sector. The educational keywords used were: "school" OR "college" OR "university" OR "education" OR "student" OR "teacher" OR "curriculum" OR "class" OR "exam" OR "homework" OR "teaching" OR "academia" OR "academic". The cleaned dataset contains 13,339 entries.

*3.3  Topic Modeling*

To address the research question, we employed Latent Dirichlet Allocation (LDA) algorithm on the tweets corpus for topic modeling. LDA is a generative probabilistic model for discrete datasets such as text corpora ([Blair, Bi, & Mulvenna, 2020](#)) and is a particularly popular method for fitting a topic model as it can be used to find natural groups of items even when it is not certain what is being searched.

---

[1] https://www.kaggle.com/datasets/konradb/chatgpt-the-tweets

Before proceeding with topic modeling algorithm. We pre-processed our dataset using the following steps: lowercasing, punctuation removal, stop words removal, identifying n-grams, and lemmatization.

Fine-tuning the hyperparameters is necessary to achieve optimal outcome. LDA hyperparameters includes the Dirichlet parameters $\alpha$ and $\beta$, and the number of topics ($K$). The higher the $\beta$, the more words the topics consist of; likewise, the higher the $\alpha$, the more diverse the topics are. To identify the most suitable values for these three hyperparameters, we performed a grid search. We exhaustively experimented with 600 different combinations of hyperparameters and the best model was evaluated and selected through comparing the topic coherence, $Cv$. A higher $Cv$ coherence value indicates greater coherence among the words within the topic, signifying a more optimal topic representation.

## 3.4 Sentiment Analysis

RoBERTa ([Liu et al., 2019](#)) is a state-of-the-art language model that has gained significant attention in the field of natural language processing. [Loureiro et al. (2022)](#) extended the original RoBERTa-base model by training it on a larger corpus of tweets. The resulting model, the TimeLMs, is particularly well-suited for Twitter data where most tweets consists of single sentences and topics of discussion change often and rapidly ([Del Tredici, Fernández, & Boleda, 2019](#)). In this paper, we employed [Loureiro et al. (2022)](#)'s TimeLMs for sentiment analysis.

Prior to performing sentiment analysis, we preprocessed the data further following the steps in Loureiro et al. (2022) by replacing user handles and URL links with generic placeholders (@user and http), except for verified users.

### 3.5 Occupation Extraction

Following the approach employed by Atalay et al. (2020), Koonchanok, Pan and Jang (2023), and Zhao, Xi and Zhang (2021). The occupation associated with each tweet is determined by comparing the user's description with a predefined list of occupation titles. We preprocessed the user's descriptions and matched each user description with job titles from O*NET, a freely available online database that includes job titles and their alternative names across the entire U.S. economy.

## 4 Results

### 4.1 Topic Modeling

The grid search results in a nine-topic LDA model with an asymmetric $\alpha$ value and a $\beta$ value of 0.91. These hyperparameter values provide the most coherent representation of the corpus with a coherence score of 0.435. Figure 3 shows that the coherence score peaks when the number of topics is set to nine.

Figure 3. Coherence scores of LDA with different number of topics ($\alpha$=asymmetric, $\beta$=0.91).

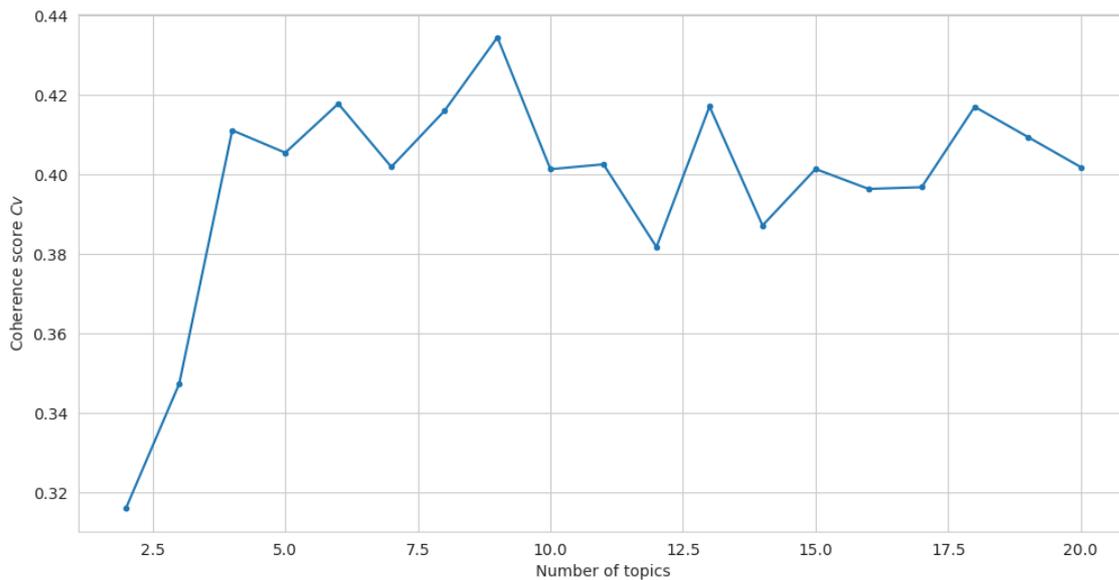

The results of LDA are shown in Table 1 and Figure 4. Our approach involved selecting representative words and conducting a thematic analysis of highly probable tweets to assign names to the topics. We identified 9 topics related to ChatGPT in education discussed by Twitter users.

*General Use in School*. A range of experiences and perspectives on ChatGPT's use in school is highlighted in this topic. The tweets reflect diverse applications of ChatGPT in education, from practical aids in lesson planning to sparking creativity and facilitating quick information retrieval for presentations. Overall, these anecdotes showcase the multifaceted ways in which ChatGPT is being utilized in school settings.

*Ability to Pass Exams*. The tweets highlight the remarkable achievements of ChatGPT in passing various exams across different domains, showcasing both its successes and limitations. The tweets highlight that ChatGPT has demonstrated its ability to excel in certain exams, particularly in law and medicine, but its limitations in specific exams emphasize the ongoing challenges in developing comprehensive AI models.

*Invitation to Join Events*. These tweets invite interested individuals to join webinars, workshops, and podcasts relevant to ChatGPT in educational settings. The content of these events mostly emphasizes the transformative potential of ChatGPT in enhancing teaching and learning experiences.

*Impact of ChatGPT*. Tweets on this topic discussed the potential impact of ChatGPT on education. Several tweets express concerns about the potential obsolescence of traditional education systems in the face of AI advancements, and questions are raised about the future aspirations of children and the need for a reformed education system. The urgency of addressing the educational crisis caused by ChatGPT is highlighted, calling for bold action and innovative solutions in higher education.

*Students' Use for Assignments*. This topic highlights a growing trend where students are leveraging ChatGPT for assistance in their academic tasks. Concerns were raised as educators and teachers are grappling with the challenges posed by ChatGPT, as it can effectively imitate undergrad-level essays, potentially impacting the evaluation process.

*Evaluating Response Accuracy*. In this topic, Twitter users typically attempt to compare and verify responses across diverse academic disciplines such as physics, mathematics, and chemistry, as well as conceptual fields like philosophy. The tweets collectively suggest that while ChatGPT demonstrates impressive capabilities, it is not infallible and has room for improvement, especially in terms of accuracy, reasoning, and avoiding the creation of misleading information.

*Academic Integrity*. Tweets on this topic provide discourse on concerns related to ChatGPT and its impact on academic integrity. Notably, Cambridge University Press has introduced guidelines aimed at maintaining transparency, preventing plagiarism, ensuring accuracy, and upholding originality in research that utilizes generative AI tools like ChatGPT. For instance, the policy requires authors to declare the use of generative AI tools. The emergence of tools like GPTZero, built by a Princeton student, underscores the ongoing battle against AI-driven plagiarism. Overall, the intersection of AI and academic integrity necessitates ongoing discussions and adaptive policies to navigate the evolving landscape of education and research.

*Anecdotal Story*. Tweets on this topic offer a diverse array of perspectives and experiences related to ChatGPT. Users share their favorite ChatGPT prompts, instances where ChatGPT provided definitions for technical terms, and situations where ChatGPT made serious accusations.

*Policy and Social Concerns*. While there is optimism about AI's ability to revolutionize education, concerns persist regarding issues such as plagiarism, cheating, and compromised writing skills. The ongoing discourse reflects a dynamic tension between the potential benefits and ethical considerations associated with the integration of ChatGPT in educational settings.

Figure 4. LDA results topics and keywords. Weights correspond to the word's importance in a particular topic. Word counts indicate the exact word frequency.

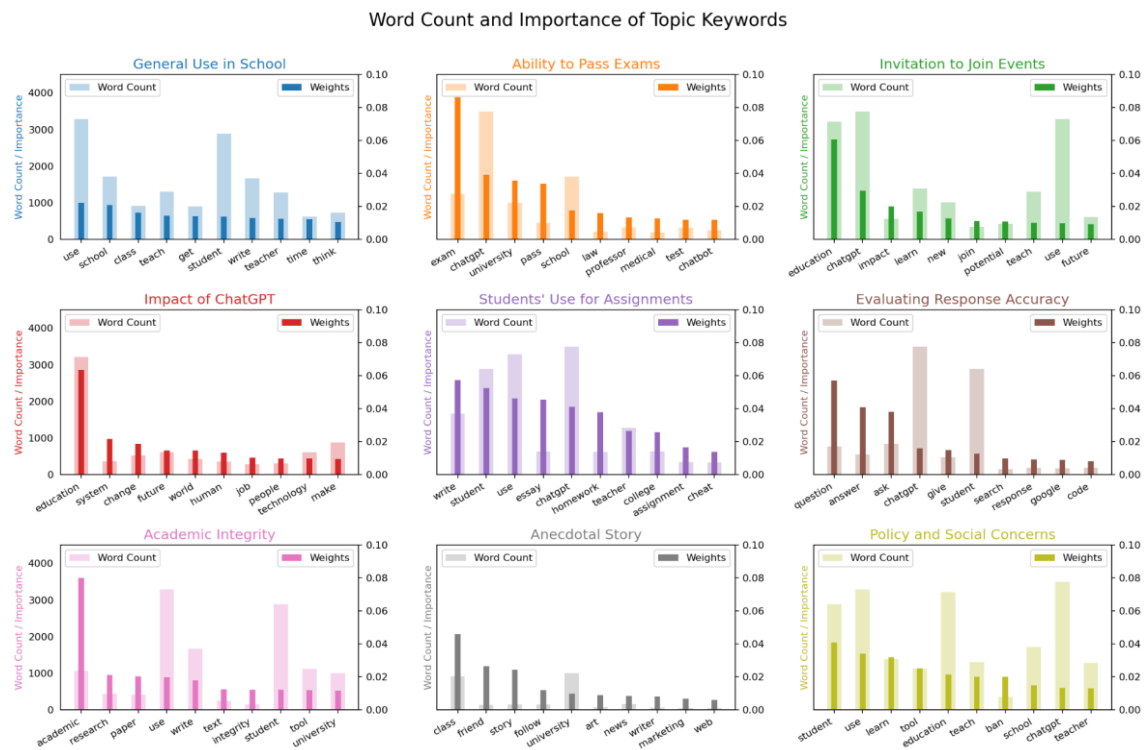

Table 1. LDA results topics and tweet examples.

| Topic | Count (%) | Representative Tweet |
| --- | --- | --- |
| General Use in School | 3346 (25.08%) | ChatGPT is the perfect tool to work smarter, not harder: Prompt example - "Make an advanced C1-level English vocabulary matching exercise with 15 useful words for the topic of Generation Gap." - Done in a minute. Check, edit, use in-class. |
| Ability to Pass Exams | 1079 (8.09%) | ChatGPT Just Passed The Bar and Medical School Exams. The ChatGPT-3 powered chatbot passed the final exam of the MBA course with flying colors & the performance of ChatGPT has scored between a B- and B grade. Microsoft invested $10b in OpenAI. Fad or trend's here to stay? |
| Invitation to Join Events | 2705 (20.28%) | Looking to level up your teaching game? Join us on June 13th for a virtual day of learning focused on AI and ChatGPT in edu. Whether you're a beginner or advanced, our sessions will equip you with valuable insights & practical strategies. |
| Impact of ChatGPT | 1352 (10.14%) | Because there may be social transformations that are happening too fast, that are going to leave people jobless and in turmoil. We need to change our education system, our social welfare system, and make sure people can shift easily to other jobs. |
| Students' Use for Assignments | 1505 (11.28%) | A college student got a top grade for an essay written with the help of ChatGPT. The student said that he submitted a 2500-word essay written by ChatGPT and another without the help from the bot. For essay that he wrote himself he got a 2.1 low grade. However for the AI-assisted essay he was given a first-Highest Grade he'd be awarded. He also notified that he did not copy everything word by word but he would prompt questions to get access to information much quicker. |
| Evaluating Response Accuracy | 823 (6.17%) | I tested the ChatGPT capacity for writing utility functions for basic image analysis. These are the type of tasks I give to an undergrad or first-year PhD student in materials sciences. It made some of the same mistakes a student usually does, but overall did pretty good. |
| Academic Integrity | 845 (6.33%) | There seems to be a lot of anxieties surrounding academic integrity due to ChatGPT, however, issues regarding academic integrity have existed long before AI in Education. Fighting and resisting the tool are not the solution, challenging our current assessment design is. |
| Anecdotal Story | 250 (1.87%) | Yikes. ChatGPT says I'm a former music librarian. NOT YET I'M NOT! I'm a hybrid librarian/postdoc researcher, but not a former anything (Discounting former shop assistant, barmaid, farm worker, ballet class accompanist ...). |
| Policy and Social Concerns | 1434 (10.75%) | New York City's Department of Education has restricted ChatGPT access across its public school devices and networks. The move could usher in similar restrictions across other school systems and educational institutions. Access restricted following requests from schools. |

In addition, we investigated the chronological patterns of discussions related to ChatGPT in the context of education. Figure 5 depicts the timeline of tweets associated with each topic. In general, the patterns for all topics tend to increase or decrease simultaneously. Initially, there was a notable surge in tweets concerning ChatGPT in education, particularly the "General Use in School" and "Students' Use for Assignments." Across all topics, this initial uptick was followed by a decline and subsequent surge in January. This surge can be attributed to ChatGPT passing MBA and law exams, coupled with the prohibition of ChatGPT usage in New York City schools (Rosenblatt, 2023). Subsequently, the trends gradually decreased, only to rise again in March, likely triggered by the release of OpenAI's highly anticipated GPT-4 model (Weitzman, 2023), as well as the call for a pause on AI development by tech leaders (Future of Life Institute, 2023).

Figure 5. The trends of each topic.

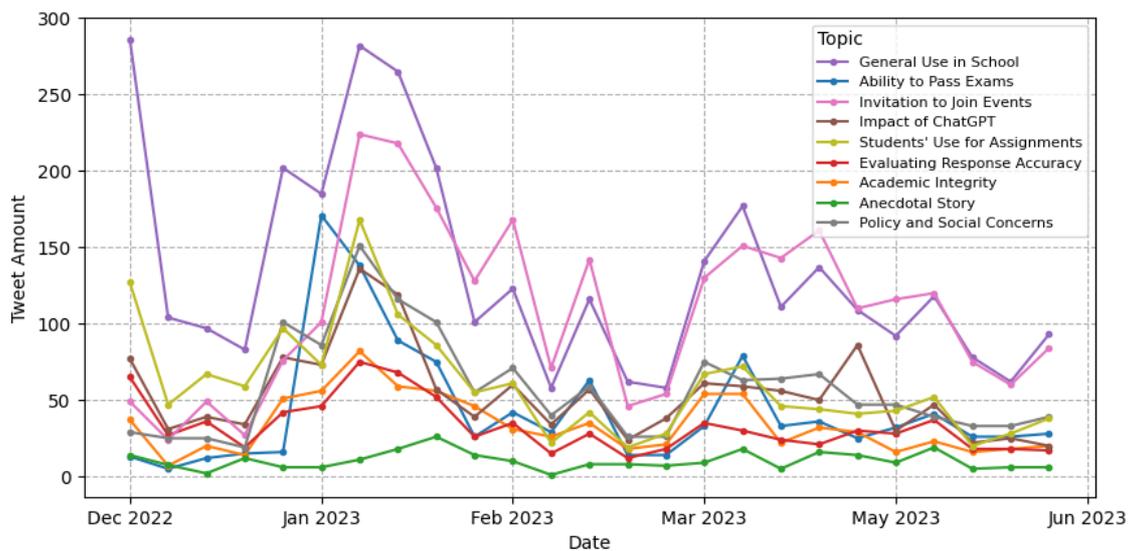

Figure 6 illustrates the t-distributed stochastic neighbor embedding (t-SNE) representation of the LDA outcomes. t-SNE is an algorithm for reducing dimensionality by presenting high-dimensional data in two dimensions while retaining the most significant structure. Figure 6 shows that the nine topics were well separated in the two-dimensional t-SNE visualization.

Figure 6. t-SNE visualization of LDA results.

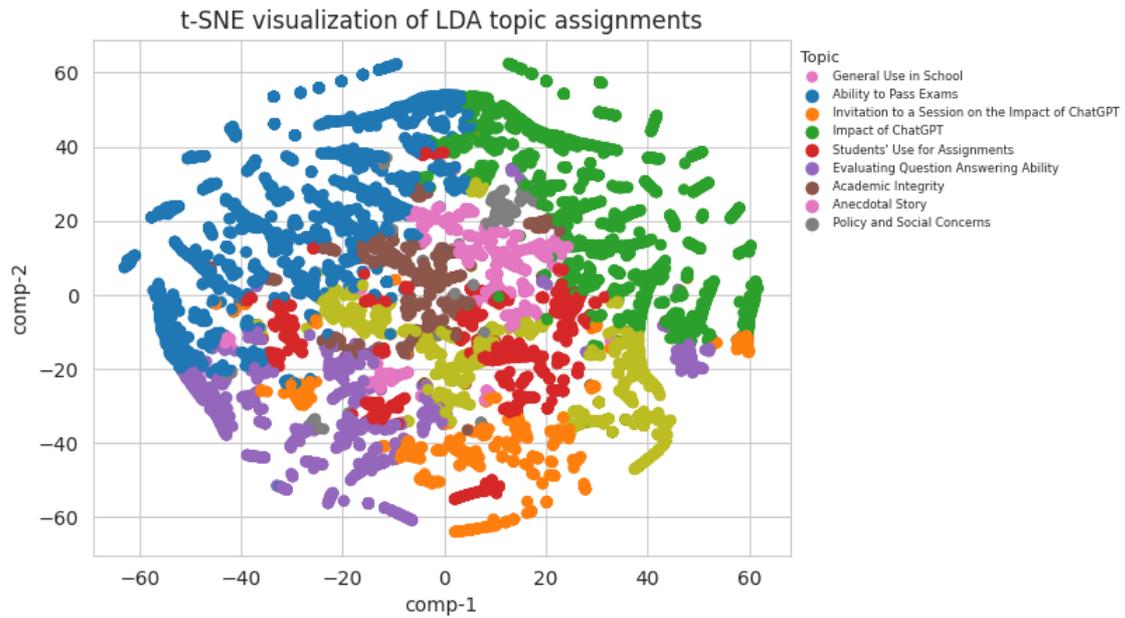

## 4.2 Sentiment Analysis

Figure 7 shows that the RoBERTa-based TimeLMs sentiment model classifies most tweets as positive and neutral (42.2% and 44.4%, respectively). Only a small portion (13.4%) were categorized as negative. These results show that people had a non-negative outlook toward ChatGPT in education. As shown in Figure 8, the positive sentiment keywords indicate that the majority of positive tweets center around discussions regarding the potential enhancement of the learning and teaching environment through the utilization of ChatGPT. In contrast, negative tweets predominantly express concerns related to academic misconduct, particularly the use of ChatGPT for completing homework assignments.

Figure 7. Distribution of positive, neutral, and negative sentiments of the tweets.

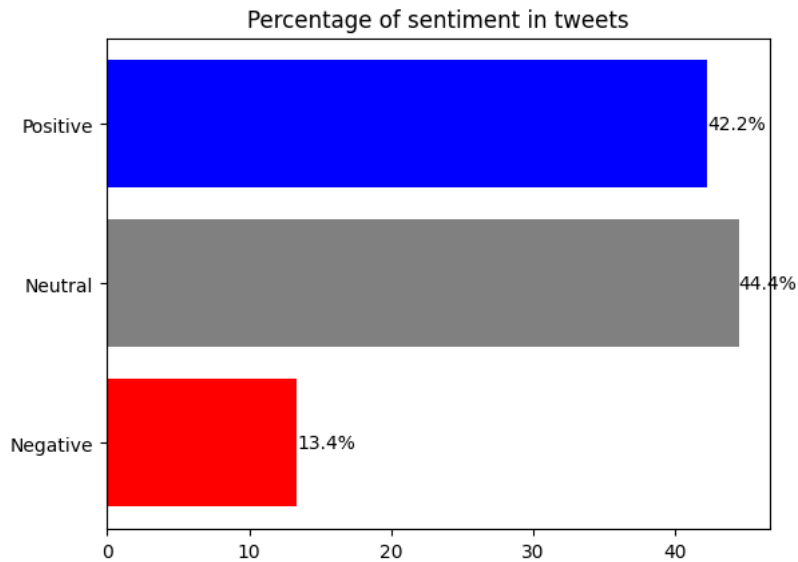

Figure 8. Word clouds of frequent positive (A) and negative (B) keywords related to ChatGPT in education.

A                                              B

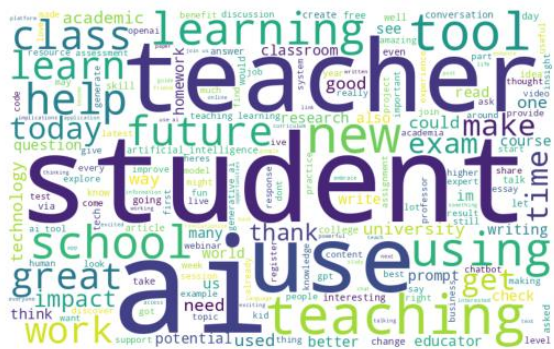 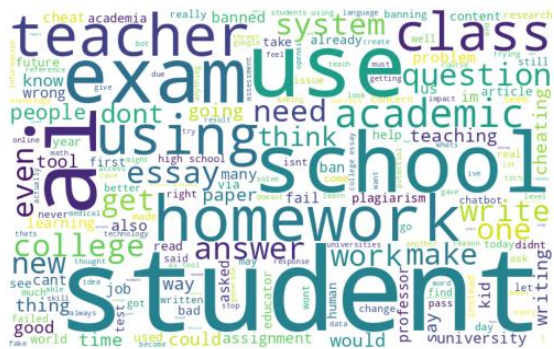

The sentiment trend during the study period is depicted in Figure 9. The sentiment analysis indicates an overall positive outlook among Twitter users regarding the application of ChatGPT in education, as evidenced by the consistent position of the green line above the blue line. Figure 9 also illustrates a significant upsurge in both positive and negative sentiments in January and March. This pattern corresponds to the trends depicted in Figure 5.

Figure 9. Sentiment analysis of positive, neutral, and negative tweets related to ChatGPT in education.

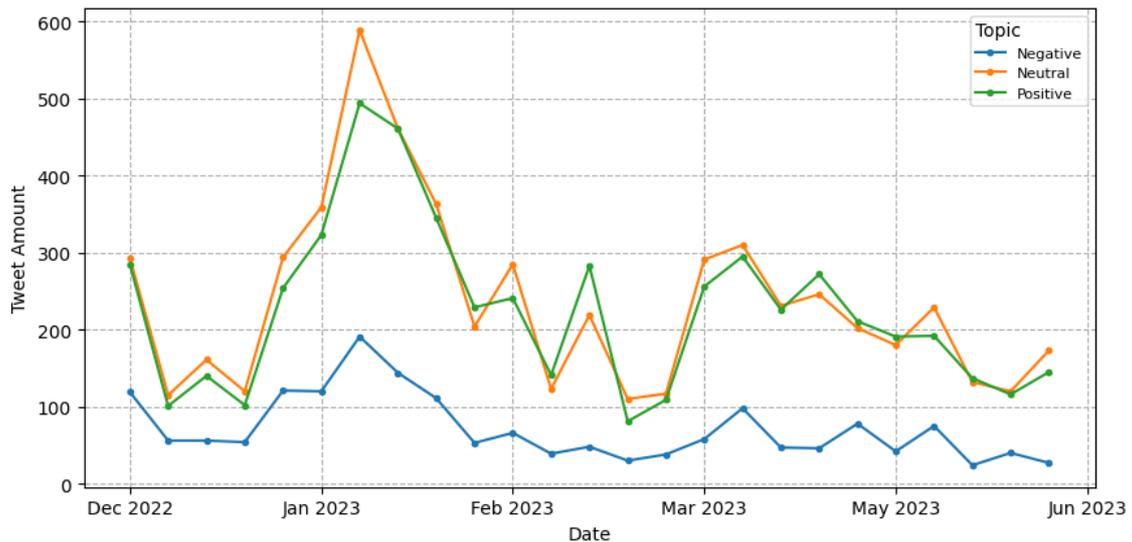

In Figure 10, we present the sentiment distribution across topics. The distribution reveals a prevalence of positive tweets across most topics. Notably, the topic "Invitation to Join Events" exhibits the highest proportion of positive tweets, with only 1.15% of tweets classified as negative. This is expected as individuals inviting others to participate in an event would typically convey their messages with a positive tone. The topics "General Use in School" and "Policy and Social Concerns" rank second and third, respectively. This further emphasizes a positive viewpoint regarding the application of ChatGPT in educational settings. In contrast, the topic "Students' Use for Assignment" displays the highest proportion of negative tweets, accompanied by the lowest proportion of positive tweets among all topics. This suggests that while individuals generally endorse the utilization of ChatGPT in educational settings, using the tool for academic dishonesty, such as cheating on assignments, is deemed unacceptable. This observation is reinforced by the ratio of positive to negative tweets concerning the topic "Academic Integrity." Another notable topic with a relatively high proportion of negative tweets is "Evaluating Response Accuracy," highlighting instances where ChatGPT may produce incorrect, inaccurate, or incomplete information.

Figure 10. Sentiment distribution per topic.

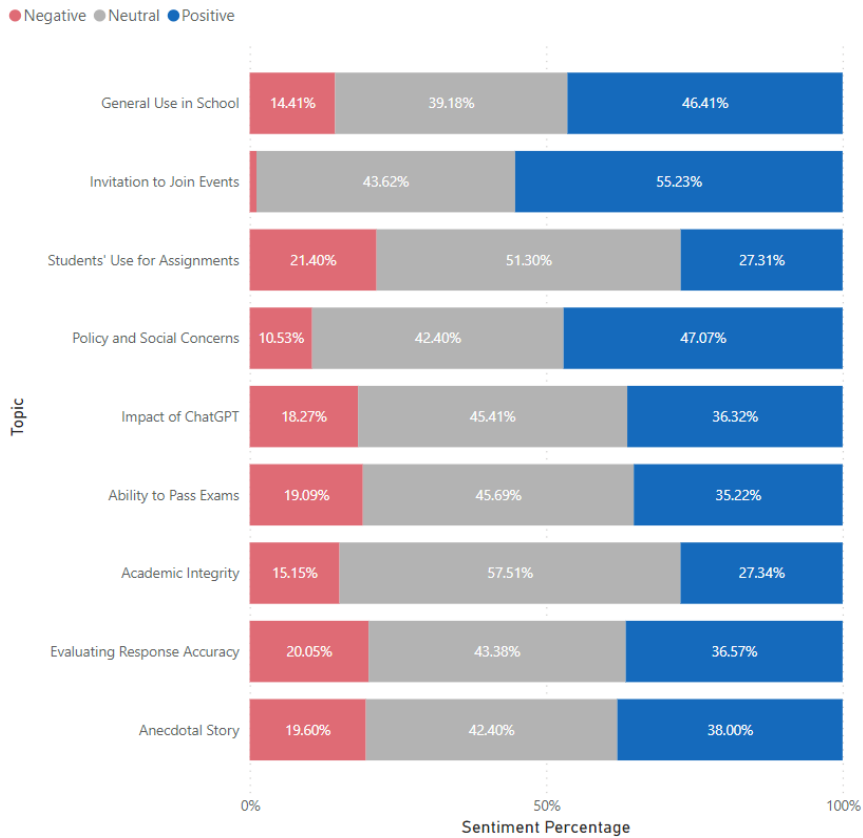

## 4.3 Occupational Differences

A breakdown of occupation frequencies is provided in Table 2. Table 2 shows that "Educational Instruction and Library" occupations are prevalent, comprising 15.52% of instances, followed by "Arts, Design, Entertainment, Sports, and Media" occupations at 7.69%. "Production" and "Management" occupations contribute 6.28% and 5.23%, respectively. The remaining major occupation groups each comprise less than 5% of the dataset.

Table 2. Occupation distribution.

| Occupation | Count | Percentage |
|---|---|---|
| Educational Instruction and Library | 2,070 | 15.52% |
| Arts, Design, Entertainment, Sports, and Media | 1,026 | 7.69% |
| Production | 838 | 6.28% |
| Management | 698 | 5.23% |
| Life, Physical, and Social Science | 481 | 3.61% |
| Computer and Mathematical | 463 | 3.47% |
| Healthcare Practitioners and Technical | 403 | 3.02% |
| Personal Care and Service | 350 | 2.62% |
| Others | 1,869 | 14.01% |
| Undefined | 5,141 | 38.54% |

Others refers to occupations with frequency less than 2%. Undefined refers to tweets that do not contain a user description that aligns with O*NET job titles.

Figure 11 illustrates the interrelation between tweet topics and major occupational groups. Predominantly, "Educational Instruction and Library" occupations emerge as the dominant occupation group across various subjects, notably claiming a significant portion in "General Use in School," "Invitation to Join Events," and "Policy and Social Concerns," with topic percentage contributions of 33.89%, 38.05%, and 40.25%, respectively. This outcome is expected as the dataset is specifically filtered for ChatGPT in educational contexts. It is noteworthy that "Healthcare Practitioners and Technical" occupations are not prominent across all topics, except the topic "Ability to Pass Exams," possibly stemming from the surge of attention when ChatGPT's capability to pass the US Medical Licensing Exam in January 2023. Similarly, the proportion of "Life, Physical, and Social Science" occupations are relatively small, except for the topic "Academic Integrity." This observation is reasonable considering that individuals within this occupational group are scientists and social scientists who often engage in diverse academic activities.

Figure 11. Occupation distribution per topic.

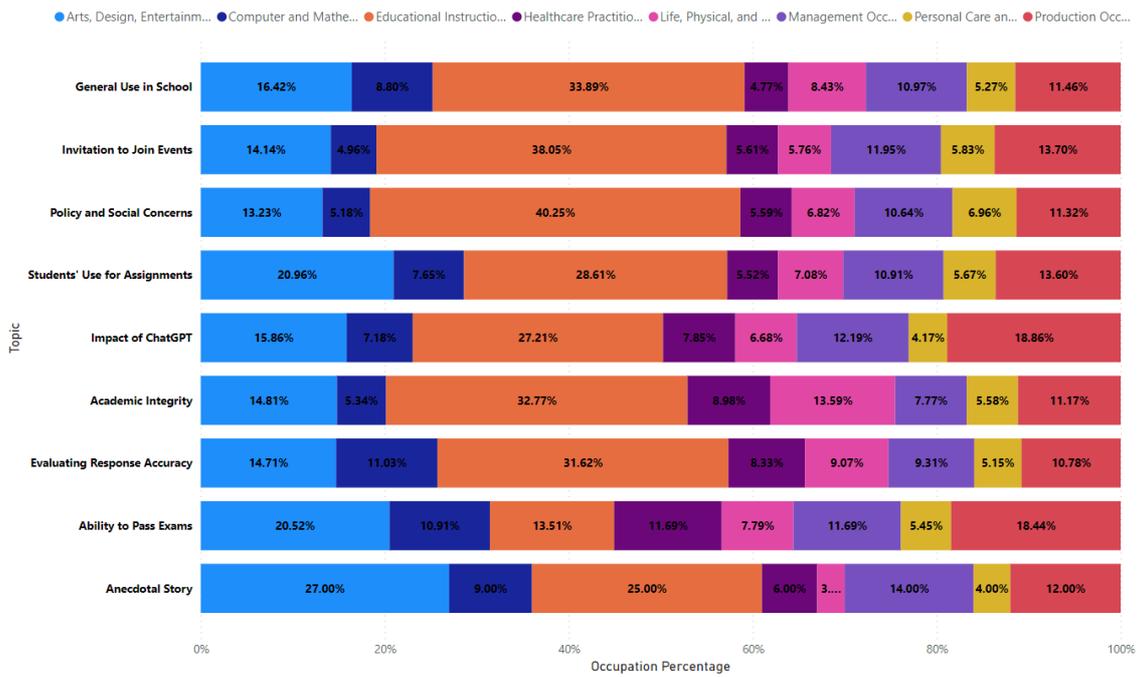

Figure 12 displays the sentiment across various major occupational groups. The "Educational Instruction and Library," "Management," and "Computer and Mathematical" occupations have the highest proportion of positive sentiments (49.08%, 46.42%, and 46.00%, respectively). On the other hand, "Life, Physical, and Social Science" and "Arts, Design, Entertainment, Sports, and Media" occupations have the highest proportion of negative sentiment (17.46% and 16.37%, respectively). This is in line with previous observation that ChatGPT's potential to assist in instructional activities and enhance educational materials is deemed positive. Moreover, given the technical nature of "Computer and Mathematical" occupations, it is expected that professionals in these fields would appreciate ChatGPT's groundbreaking capability. On the other hand, professionals in "Life, Physical, and Social Science" may harbor reservations regarding ChatGPT possibly due to concerns about academic misconduct, as highlighted previously. "Arts, Design, Entertainment, Sports, and Media" professionals may express negativity towards ChatGPT possibly reflecting concerns about authorship, authenticity, and job displacement.

Figure 12. Sentiment distribution per occupation.

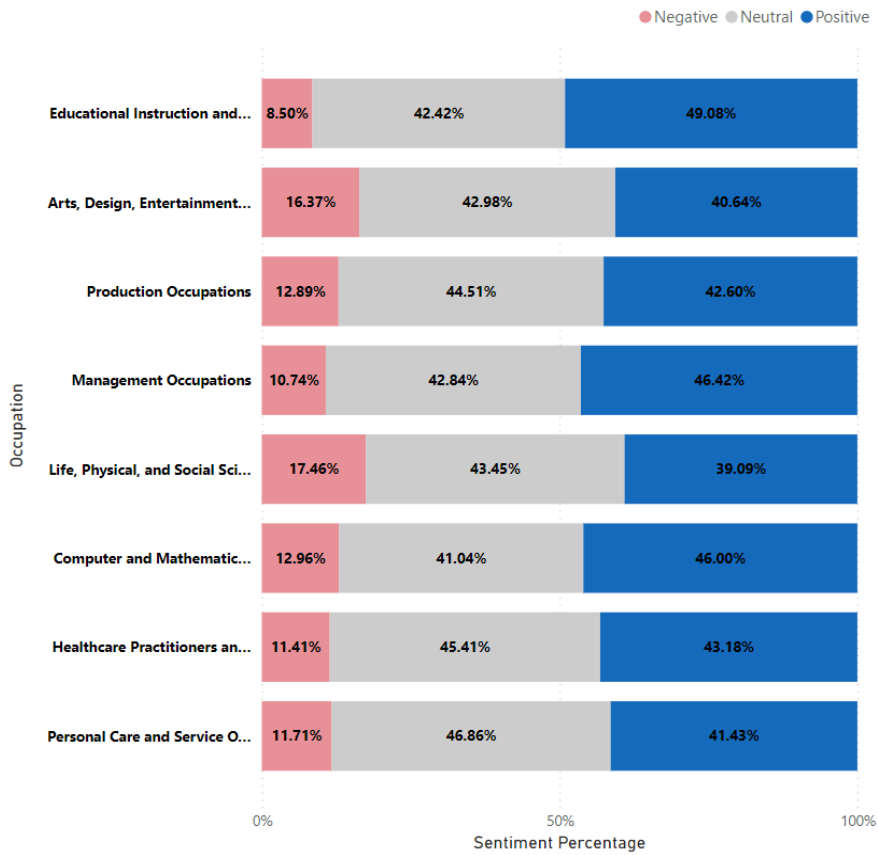

## 5 Conclusion

This study analyzed Twitter data to explore ChatGPT's use in education, using topic modeling and sentiment analysis to identify key themes and sentiments. It also examined occupational differences in perceptions, revealing diverse public attitudes towards ChatGPT in education.

## 6 Declaration of Interest Statement

The authors declare that there is no conflict of interest.

# 7 References

**Uncategorized References**